\documentclass[10pt,aps,prb,groupedaddress,superscriptaddress,showpacs,twocolumn]{revtex4-1}

\usepackage{amssymb}
\usepackage{amsmath}
\usepackage{latexsym}
\usepackage{revsymb}
\usepackage{graphicx}

\begin{document}
\def\Z{\hbox{{\sf Z}\kern-0.4em {\sf Z}}}

\title{
Nonlocal Edge State Transport in Topological Insulators
}

\author{Alexander P. Protogenov}
\email{alprot@appl.sci-nnov.ru} 
 \affiliation{Institute of Applied Physics of the RAS, Nizhny Novgorod 603950, Russia}
 \affiliation{ Donostia International Physics Center (DIPC), 20018 San Sebasti\'an/Donostia, Basque Country, Spain}

\author{Valery A. Verbus}
 \affiliation{Institute for Microstructures of the RAS, Nizhny Novgorod 603950,
Russia}

\affiliation{NRU-Higher School of Economics, Nizhny Novgorod 603950,
Russia}

\author{Evgueni V. Chulkov}
\affiliation{ Donostia International Physics Center
(DIPC), 20018 San Sebasti\'an/Donostia, Basque Country, Spain} 
\affiliation{Departamento de F\'isica de Materiales EHU, Centro de
F\'isica de Materiales CFM-MPC and \\Centro Mixto CSIC-UPV/EHU,
20018 San Sebasti\'an/Donostia, Basque Country, Spain}
 \affiliation{Tomsk State University, Tomsk 634050, Russia}

\begin{abstract}

We use the $N$-terminal scheme for studying the edge state transport
in two-dimensional topological insulators. We find the
universal nonlocal response in the ballistic transport approach.
This macroscopic exhibition of the topological order offers different  areas for applications.

\end{abstract}
\pacs{73.63.-b, 73.20.-r, 79.60.Bm}
\maketitle

\section{Introduction}

A topological insulator is a quantum phase of matter with gapless
electron states on the surface and gapped ones, in the bulk. In
two-dimensional systems, conducting electron states propagate along
the edge of the topological insulator. These states have a linear
Dirac dispersion and their spin is locked to the momentum
\cite{HK,QZ,M}. The edge or the surface electron states are
topologically protected. The time inversion symmetry leads to the
topological protection of each Kramers partner and as result to
suppression of the backscattering. Therefore, the topological
protection against moderate structural disorder is the base for many
technological applications of topological insulators.

The features of topological insulators are presented in nano-scale samples.
Actually, the topological order in topological insulators belongs to the phase state
which is called a short-range order created by the entangled quantum states \cite{CGW}.
By this term we mean that this phase state is formed on the scale of the order of the
lattice constant. The topological phase, which is formed on larger scales -- so-called
long-range topological order -- belongs to a fundamentally different class \cite{CGW}
of the entangled quantum states. One of its features is nonlocality in the Hilbert space.
The fact is that the total quantum state is not equal to the product of the local electron states.
In other words, there are long-range correlations of topological excitations in this state.
Usually, the topological order affects the collision frequency \cite{CHSS}.
The geometrical Berry phase of quantum states in topological insulators resulting in
interference phenomena leads to the weak antilocalization effect \cite{BM}.

In this paper we consider the universality of the transport characteristics in two-dimensional
topological insulators. They exhibit nonlocal topological order.
The existence of nonlocality in the 3D case follows
already from the representation \cite{MRW} of the topological invariant.
The nontrivial value of this Hopf invariant, being equal to unity, means that the
two loops are linked. We focus on the significantly nonlocal response which was found in the recent
experiments \cite{RBBMMQZ,BKTGFX}.

\section{Nonlocal response}

The features of electron states in two-dimensional topological
insulators are exhibited in the quantum spin Hall effect
\cite{KM,BHZ,KWB,KBM}. When studying the degree of nonlocality, we
use the approach of Refs. \cite{RBBMMQZ,TH}. The main distinction
here is that we employ the general $N$-terminal scheme. It is
convenient to use the ballistic Landauer-B\"utticker approach
\cite{B} writing the current $I_{i}$, injected through the terminal
$i$ into a sample as
\begin{equation}
\label{eq0}
I_{i}=\frac{e^2}{h}\sum\limits_{j=1}^{N}(T_{ji}V_{i} - T_{ij}V_{j}) \, \, .
\end{equation}
Here, $V_{j}$ is the voltage on the terminal $j$, $e$ is the electron charge,
$h$ is the Planck constant, and $h/e^{2}$ is the resistance unit. We will equate it further
to unity. $T_{ij}$ is the matrix element of transmission from the terminal $i$ to
the terminal $j$, and $N$ is the total number of terminals.

The edge electron modes in topological insulators propagate in two directions.
We will put that the transmission coefficients between neighboring terminals are equal
to unity: $T_{i+1,i}=T_{i,i+1}=1$ and other probabilities are equal to zero.
The $N$-terminal scheme implies the use of the periodic boundary conditions
$T_{N+1,N}=T_{1,N}$, $T_{N,N+1}=T_{N,1}$ in the both directions of the
edge state propagation.

Let us consider the number of terminals $N$ as a tuning parameter. The labels of terminals,
whose voltages we will measure, will contain
information about the influence of the edge state current between the terminals,
through which the current flows about the voltage distribution on other terminals.
This distribution will define the degree of the response nonlocality.
For example, in the case of $N=4$ [14], if the current flows from the terminal
$1$ to the terminal $4$, we may be interested in the resistance between these terminals and
the nonlocal response, expressed in terms of the resistance between
terminals $1$ and $2$, $2$ and $3$, and also $1$ and $3$. It is evident that with
the increase in the number of terminals $N$, the resistance between terminals
$1$ and $N$, through which the current flows, will tend to unity. As regards the
adjacent terminals such as $1$ and $2$, the measured resistance will tend to zero according to a certain law decreasing with the increasing number $N$.

Let the current $I_{1N}$ flows through terminals $1$ and $N$ (see Fig. 1)
and the voltage be measured on all other terminals.
In this case, the equation which determines the voltage on the contacts, has the form
\begin{equation}
\label{eq2}
A\,V = I \, \, ,
\end{equation}
where the matrix $A$ equals $A_{ij}=2\delta_{ij} - \delta_{i,j+1} -
\delta_{i,j-1} - \delta_{i,1}\,\delta_{j,N} - \delta_{i,N}\,\delta_{j,1}$, $\delta_{ij}$
is the Kronecker delta,
$1 \leq i,j \leq N$, $V=(V_{1}, V_{2}, \dots V_{N-1},V_{N})$, and
$I=I_{1N}(1,0, \dots 0,-1)$.
We are interested in the difference
between the voltage on terminals. Since the vector $V$ is invariant with respect
to the constant value shift, we may assume that $V_{N}=0$.

For arbitrary $N$, the solution of Eq.(\ref{eq2}) has the form
$V_{i} = I_{1N}\left( 1 - \frac{i}{N}\right)$. Therefore, the resistance $(V_{1}-V_{N})/I_{1N}$ between terminals $1$ and $N$ equals $R_{1N,1N}=(N-1)/N$. The nonlocal resistance
$R_{kl,ij=}(V_{i}-V_{j})/I_{kl}$ at $k=1, l=N$ when measuring the voltage
between the terminals $i$ and $j$ is
\begin{equation}
\label{eq3}
R_{1N,ij} = \frac{j-i}{N} \, \, .
\end{equation}

\begin{figure}
\includegraphics[width=\columnwidth]{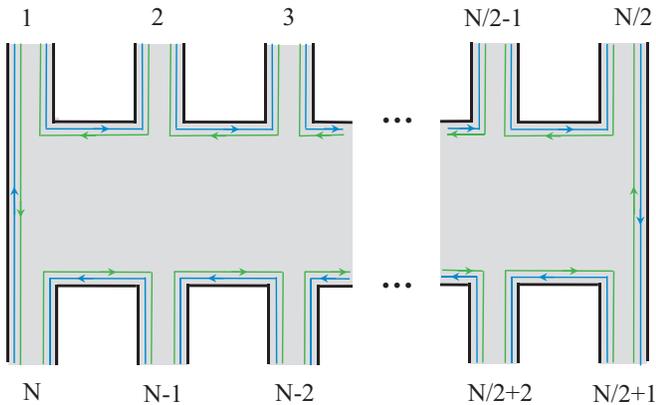}
  \caption{Distribution of edge states in a two-dimensional $N$-terminal insulator.}
  \label{fig:BasisTransform}
\end{figure}

In Eq.(\ref{eq2}), we implied that the current was conducted through the first and the last terminals. Thus, Eq. (\ref{eq3}) is valid only in this case. We may also conduct the current through
terminals $1,4$ or $1,3$, for example, in such a way as it was realized in the
experiment \cite{RBBMMQZ,BKTGFX} at $N=6$. To find the voltage distribution in the situation, when
the current flows from the terminal $1$ to the terminal $k$, in the right-hand
side of Eq.(\ref{eq2},) we have to use the equation for the current in the form
of $I=I_{1k}(1,0,\dots -1,\dots 0)$. Here, $-1$ is located in the $k$ place.
The exact solution of Eq.(\ref{eq2}) in this general case has the form
\begin{equation}
\label{eq4}
V_{i} = I_{1k}\left( 1 - \frac{i}{N}(1-k+N)\right) \, \, ,
\end{equation}
if $1 \leq i \leq k$, and
\begin{equation}
\label{eq5}
V_{i} = I_{1k}(1-\frac{i}{N})(1-k) \, \, ,
\end{equation}
if $k \leq i \leq N$.

The values of the resistance $R_{14,14}=3/2$ and $R_{14,23}=1/2$ calculated with the use
of Eq. (\ref{eq4}) coincide with the experimental results \cite{RBBMMQZ}.
When the current flows from terminal $1$ to terminal $3$, the experimental resistance \cite{RBBMMQZ}
equals $R_{13,13}=4/3$, $R_{13,56}=1/3$. This can be obtained from Eq. (\ref{eq5}) at $k=3$ and $N=6$.

\section{Discussion}

The chiral edge states in the quantum Hall effect propagate only in
one of the two possible directions. The matrix elements $T_{ij}$ of
the transmissions for such states in Eq. (\ref{eq0}) are not equal
to zero for terminals with indices $j>i$: $T_{i,i+1}=1$, and
$T_{i+1,i}=0$. Therefore, the resistance $R_{1N,ij}$ appears to be
equal to zero. In other words, the nonlocal resistance in systems
with broken time reversal symmetry is absent. We should emphasize
that the main feature of the considered situation, which governs the
obtained results, is the existence of the time reversal symmetry.
This symmetry is manifested itself at the macroscopic level in the
form of nonlocal effects, caused by existence of the helical edge
states. The form of responses and the universal character of the
obtained numbers depend on the experimental conditions. For example,
on temperature which determines the contribution of the inelastic
backscattering processes to the conductivity. The experimental data
\cite{RBBMMQZ} show a high degree of universality. The universality
of the  $R_{1N,1N}=(N-1)/N$, $R_{1N,ij}=(i-j)/N$ and Eqs.
(\ref{eq4}), (\ref{eq5}) can be verified in the experimental setups
appropriate for study the quantum spin Hall effect by varying the
total number of terminals and indices of current-carrying terminals.

If the time reversal symmetry is broken by magnetic impurities, and the condition of the
absence of the backscattering is weakened, the matrix elements $T_{i,j}$
can be written as $T_{i+1,i}=1+k_{1}$, $T_{i,i+1}=k_{2}$ \cite{WLZZ}.
Here $k_{1} < 1$ and $k_{2} < 1$ are constants, and the unity in the matrix element $T_{i+1,i}$
means the existence of the chiral edge mode. In this quasi-helical edge state \cite{WLZZ},
the voltage distribution on the terminals, when the current flows through the voltage leads $1, N$, has the form
\begin{equation}
\label{eq6}
V_{i} = \left(\frac{1-r^{N-i}}{1-r^{N}}\right)\frac{I_{1N}}{1+k_{1}} \, \, ,
\end{equation}
where $r=k_{2}/(1+k_{1})$.

In this paper, we focus on the universal exhibition of topological order in the transport properties of ideal two-dimensional topological insulators in the most straightforward and representative form. The study of the transport characteristics of the so-called ideal topological insulator $SmB_{6}$ revealed \cite{BKTGFX} that in the three-dimensional case  the transport properties significantly depend on the geometry of samples and terminal assignments. A deviation from the universal behavior takes place also in two-dimensional systems. It occurs due to metal droplets inside real contacts. This phenomenon can be described in terms of an additional terminal. The effect of this and other factors such as a finite width of the terminal, reflections from the internal interfaces, and other conditions on amplitudes of the transitions between current and voltage terminals has been studied in Ref. \cite{BRBHMMQZ}.

Let us clarify the role played by the contacts in the edge state transport following the approach and the main conclusions of Refs. \cite{RBBMMQZ,BRBHMMQZ,BRNKBHHSM} in more detail. First of all, we note that a contact is not a time-reversal symmetry breaking potential that mixes counter-propagating edge states with opposite spins. Contacts are finally an electron degrees of freedom reservoir which incoherently
populates both edge state channels. An ideal contact populates both edge state channels with equal weight by injecting spin up and spin down electrons with equal probability \cite{RBBMMQZ}. This is an origin of the additional resistance produced by the contacts. A contribution of such a dephasing reservoir into an additional longitudinal resistance can be negligible under the condition $L < L_c \sim 1/\eta$ . Here $L$ is the characteristic linear size of a contact, $L_c$ is the dephasing length, and $\eta$ is the dephasing strength of the self-energy part \cite{RBBMMQZ}. Note that the self-energy should not break
the time-reversal symmetry. Decoherent behavior arises due to the existence of the dephasing reservoir with the distribution function included in the so-called lesser self-energy of the leads \cite{RBBMMQZ}.

However, there are sharp dips in the conductance even for small values of $\eta L$. They can be strong enough to completely block coherent transport at one of the edges \cite{RBBMMQZ}. Therefore, even a small dephasing region can equally affect a probe terminal. The experimental value of the maximal resistance for the six-terminal device is 1/7 instead of the theoretical prediction 1/6. Such a result is consistent with the existence of the additional dephasing region. Dephasing regions can also exist due to an inhomogeneity of the sample. The experimental results have shown \cite{RBBMMQZ} that a change in the gate voltage also affects the homogeneity of the device due to trap state charging at the semiconductor-insulator interface. This leads to an inhomogeneous potential in the gated area and to the creation of the metallic islands which exist when most of the gated regions are insulating ones. In other words, a metallic island can lead to the effect similar to an additional probe.
The experimental situations when the coherent transport is observable in the whole sample is discussed in Refs. \cite{RBBMMQZ,BRBHMMQZ}.

There are two different methods to suppress the nonlocal transport. The first approach is to make the device scale sufficiently small
so as to induce backscattering in the channels of the edge states. Backscattering occurs when the wave functions for opposite spin orientations overlape \cite{ZLCSN}. This happens for a device width about 200 nm. Therefore, if the width $W_1$ of central device strip is rather large, the deviation $T_{1N}^{'}$ from the ideal value $T_{1N}=1$ is negligibly small \cite{BRBHMMQZ}. The same condition $W > W_2$ for the absence of the tunneling
between the edges of the individual terminal is valid for the terminal width $W_2$. Measurements of the nonlocal resistance \cite{BRBHMMQZ} in devices when they are in the quantum spin Hall effect states show the values expected for the non-perturbative nonlocal edge state transport. The numerical
simulations of the scattering matrix at the metal-topological insulator interface has confirmed the negligibly small value of  $T_{1N}^{'}$ for
the employed samples. The second method to suppress the edge state contribution to nonlocal transport is to choose such nonlocal configurations that implies the edge channel transport over distances longer than the inelastic length \cite{BRNKBHHSM}. This means that the maximal number of terminals $N < N_c =
L_1/(W_2 + L_2)$ can be roughly estimated as 10 for the real experimental parameters. Here $L_1$ is characteristic sample size and $L_2$ is the distance between terminals.

Let us consider the regime of the quantum spin Hall state when the renormalized voltage $V_{g}^{*}=0$ and the resistance is maximal. The Fermi level in the bulk is located now at the center of the energy gap. In this case, the assignments of the current and voltage terminals determine the following values of the resistance
\begin{equation}
\label{eq7}
R_{1k,ij}=
\begin{cases}
\frac{j-i}{N}\left(1-k+N\right), \,\,\,\,   1 \leq i,j \leq k,  \\
\frac{j-i}{N}\left(1-k\right), \,\,\,\,\,\,\,\,\,\,\,\,\,\, \,\,\,  k \leq i,j \leq N, \\
\frac{j-i}{N}\left(1-k\right) + (k-i),\,\,  1 \leq i \leq k, \,\, k \leq j \leq N \ .
\end{cases}
\end{equation}
One can easy verify that after the interchange of the current $(1k)$ and voltage $(ij)$ probe indices and the shifts $k \to j-i+1, j \to k-i+1, i \to N-i+2$
these equations satisfy the Onsager-Casimir symmetry relations $R_{mn,kl}=R_{kl,mn}$ \cite{Bu} for the nonlocal resistances $R_{mn,kl}$. We would also like to mention the fact that topological universality of the ballistic transport due to the edge states under ideal conditions is determined by the topological properties of the electron quantum bulk states. Therefore, the considered phenomenon in a trivial insulator is absent.

\section{Conclusion}

We have described here the universal distribution of the resistances
studying the edge state transport in the two-dimensional topological
insulators in the ballistic transport regime. It is of interest to
extend the problem of macroscopic manifestation of the topological
order to similar topologically ordered systems. Another way of
copying the letter $H$ in the vertical direction as it is done in
the ladder-like structures \cite{GTKLTW} corresponds to the
four-terminal case. This is because the distribution of edge degrees
of freedom will be equivalent to the distribution taking place in
the system with one letter $H$. For modeling of the distribution of
the degrees of freedom in the systems with a long-range topological
order, we have to use $Y$-shape contacts as building blocks.

\section*{Acknowledgements}

The authors are grateful to V. Ya. Demikhovskii, G. M. Fraiman, A.
S. Sergeev for discussions and to C. L. Kane, M. B\"uttiker and S.
V. Eremeev for useful remarks and suggestions. This work was
supported in part by the RFBR Grant No. 13-02-12110.

\end{document}